\documentclass{DISproc}

\usepackage{epsfig}
\usepackage{graphicx}

\begin{document}
\title{A new approach to gluon-quark multiplicity ratio}

\author{{\slshape Paolo Bolzoni$^1$}\\[1ex]
$^1$II. Institut f\"ur Theoretische Physik, Luruper Chaussee 149, 22761 Hamburg, Germany}

\contribID{xy}

\doi  

\maketitle

\begin{abstract}
We present a new approach in considering and including both the
perturbative and the nonperturbative contributions to the 
multiplicity ratio $r$ of gluon and quark jets. The new method is motivated
by recent developments in timelike small-x resummation obtained in the 
$\overline{\rm MS}$ factorization scheme. A global analysis to fit the 
available data is also presented.
\end{abstract}

\section{Introduction}

The gluon-quark multiplicity ratio is defined as $r=N_g/N_q$, where $N_{g(q)}$ is
the number of hadrons produced in a gluon(quark) jet.
A purely perturbative and analytical prediction has been achieved by a solution to the equations for the generating functionals in the modified 
leading logarithmic approximation (MLLA)  
in Ref.\cite{Capella:1999ms} up to the so called N$^3$LOr in the expansion parameter 
$\gamma_0=\sqrt{2N_c \alpha_s/\pi}$ \,\emph{i.e.} $\gamma_0^3$.
Here the theoretical prediction is about 10\% higher than the data at the
scale of the $Z^0$ vector boson and the difference with the data becomes 
even larger at lower scales.
Among the many attempts to predict $r$ numerically, the most successfull refers to numerical sulutions to the coupled system of equations of the generating functionals 
for the quark ($Z_F$) and the gluon ($Z_G$) in the MLLA framework 
(see \emph{e.g.} \cite{Dokshitzer:1991wu}).
These numerical solutions describe well the data only above
at relatively high energies \cite{Malaza:1985jd,Catani:1991pm,Lupia:1997in}.   
This shows that the slope of the multiplicity ratio predicted
by this approach tends to be smaller than its experimental value.
An alternative approach was given in Ref.
\cite{Eden:1998ig} where equations for the derivative of the ratio of the
multiplicities are obtained in the MLLA 
within the framework of the colour dipole model. 
There a constant of integration 
which encodes nonperturbative contributions is fixed by
the data. Here a new approach is presented. 

\section{The multiplicity ratio in the effective-$\omega$ approach}

We consider the standard Mellin-space moments of the coupled gluon-singlet system 
whose evolution in the scale $\mu^2$ is governed in QCD by the DGLAP equations:
\begin{equation}
\label{ap}
\mu\frac{d}{d\mu} \left(\begin{array}{l} D_s \\ D_g
\end{array}\right)=\left(\begin{array}{ll} P_{qq} & P_{gq} 
\\ P_{qg} & P_{gg}\end{array}\right)\left(\begin{array}{l} D_s \\ D_g
\end{array}\right).
\end{equation}
The timelike splitting functions $P_{ij}$ can be computed
perturbatively in the strong coupling constant:
\begin{equation}
\label{exp}
P_{ij}(\omega,\mu^2)=\left(\frac{\alpha_s(\mu^2)}{4 \pi}\right) P_{ij}^{(0)}(\omega)+
\left(\frac{\alpha_s(\mu^2)}{4 \pi}\right)^2 P_{ij}^{(1)}(\omega)+
\left(\frac{\alpha_s(\mu^2)}{4 \pi}\right)^3 P_{ij}^{(2)}(\omega)+\emph{O}(\alpha_s^4),\quad i,j=g,q,
\end{equation}
where $\omega=N-1$ with $N$ the usual Mellin conjugate variable to the
fraction of longitudinal momentum $x$.
The functions $P_{ij}^{(k)}(\omega)$ with $k=0,1,2$ appearing in Eq.(\ref{exp}) 
in the $\overline{MS}$ scheme can be found in 
Ref.\cite{Gluck:1992zx,Moch:2007tx,Almasy:2011eq} up to NNLO and in 
Ref.\cite{Vogt:2011jv} the NNLL contributions up to  $\emph{O}(\alpha_s^{16})$ 
in the same scheme. 
Fully analytical resummed results in a closed form in the $\overline{MS}$ scheme 
are known at NLL for the eigenvalues of the siglet-gluon matrix 
\cite{Vogt:2011jv,Albino:2011cm}. 

It would be desiderable to fully diagonalize Eq.(\ref{ap}). However
in general this is not possible because the contributions to the 
splitting function matrix do not commute at different orders.
One is hence enforced to write a series expansion about the 
LO which in turn can be diagonalized. Therefore we start 
choosing a basis where the LO is diagonal (see \emph{e.g.}\,\cite{Buras:1979yt}) 
with the timelike splitting function matrix taking the form:
\begin{equation}
P(\omega)=
\left(\begin{array}{ll} P_{++}(\omega) & P_{+-}(\omega) 
\\ P_{-+}(\omega) & P_{--}(\omega)\end{array}\right),
\end{equation} 
where by definition
\begin{equation}
\label{offdiag}
P_{+-}^{(0)}(\omega)=P_{-+}^{(0)}(\omega)=0,
\end{equation}
and where $P_{\pm \pm}^{(0)}(\omega)$ are the eigenvalues of the LO splitting matrix.

Now relating the $(g,s)$ basis to this new $(+,-)$ basis, we 
can decompose the singlet and the gluon fragmentation
function symbolically in the following way:
\begin{equation}
\label{decomp}
D_a(\omega,\mu^2)=D_a^+(\omega,\mu^2)+D_a^-(\omega,\mu^2); \qquad a=s,g.
\end{equation} 
According to Eq.(\ref{ap}) the plus and minus components have the form
\begin{equation}
\label{evolsol}
D_a^\pm(\omega,\mu^2)=\tilde{D}_a^\pm(\omega,\mu_0^2)
\left[\frac{\alpha_s(\mu^2)}{\alpha_s(\mu_0^2)}\right]
^{-\frac{P_{\pm\pm}^{(0)}}{2\beta_0}}\, H_{a}^\pm(\omega,\mu^2),
\end{equation}
where the normalization factors $\tilde{D}_a^\pm(\omega,\mu_0^2)$ satisfy
\begin{equation}
\label{rlo}
\tilde{D}_g^+(\omega,\mu_0^2)=-\frac{\alpha_\omega}{\epsilon_\omega}\,
\tilde{D}_s^+(\omega,\mu_0^2)\,;
\qquad
\tilde{D}_g^-(\omega,\mu_0^2)=\frac{1-\alpha_\omega}{\epsilon_\omega}\,
\tilde{D}_s^-(\omega,\mu_0^2),
\end{equation}
with 
\begin{equation}
\alpha_\omega=\frac{P_{qq}^{(0)}(\omega)-P_{++}^{(0)}(\omega)}{P_{--}^{(0)}(\omega)-P_{++}^{(0)}(\omega)},
\quad \epsilon_\omega=\frac{P_{gq}^{(0)}(\omega)}{P_{--}^{(0)}(\omega)-P_{++}^{(0)}(\omega)}.
\end{equation}
The perturbative functions $H_a^\pm(\omega,\mu^2)$ in Eq.(\ref{evolsol})
up to NNLO may be represented as
\begin{eqnarray}
\label{pertfun}
H_a^\pm(\omega,\mu^2)&=&1+\left(\frac{\alpha_s(\mu^2)}{4 \pi}\right)
\left(Z_{\pm\pm,a}^{(1)}(\omega)-Z_{\pm\mp,a}^{(1)}(\omega)\right)\nonumber\\
&&+\left(\frac{\alpha_s(\mu^2)}{4 \pi}\right)^2
\left(\tilde{Z}_{\pm\pm,a}^{(2)}(\omega)-\tilde{Z}_{\pm\mp,a}^{(2)}(\omega)\right),
\end{eqnarray}
where the functions $Z_{\pm\pm,a}^{(1)}$, $Z_{\pm\mp,a}^{(1)}$, 
$\tilde{Z}_{\pm\pm,a}^{(2)}$ and $\tilde{Z}_{\pm\mp,a}^{(2)}$ with $a=g,s$
in terms of the timelike splitting functions up to NNLO in the $(+,-)$ basis are given by:
\begin{eqnarray}
Z_{\pm\pm,s}^{(1)}(\omega)&=&Z_{\pm\pm,g}^{(1)}(\omega)=\frac{1}{2\beta_0}\left[P_{\pm\pm}^{(1)}(\omega)
-P_{\pm\pm}^{(0)}(\omega)\frac{\beta_1}{\beta_0}\right],\\
Z_{\pm\mp,s}^{(1)}(\omega)&=&\frac{P_{\pm\mp}^{(1)}(\omega)}{2\beta_0+P_{\pm\pm}^{(0)}(\omega)
-P_{\mp\mp}^{(0)}(\omega)}\,,\quad
Z_{\pm\mp,g}^{(1)}(\omega)= Z_{\pm\mp,s}^{(1)}(\omega)\,\frac{P_{qq}^{(0)}(\omega)
-P_{\mp\mp}^{(0)}(\omega)}{P_{qq}^{(0)}(\omega)-P_{\pm\pm}^{(0)}(\omega)},\nonumber
\end{eqnarray}
and 
\begin{eqnarray}
\tilde{Z}_{\pm\pm,s}^{(2)}(\omega)&=&\tilde{Z}_{\pm\pm,g}^{(2)}(\omega)=
\frac{1}{4\beta_0}\bigg[P_{\pm\pm}^{(2)}(\omega)
-\left(P_{\pm\pm}^{(1)}(\omega)-P_{\pm\pm}^{(0)}(\omega)Z_{\pm\pm,s}^{(1)}(\omega)
\right)\frac{\beta_1}{\beta_0}\nonumber\\
&&+P_{\pm\pm}^{(0)}(\omega)\left(\frac{\beta_1^2}{\beta_0^2}
-\frac{\beta_2}{\beta_0}\right)-\sum_{i=\pm}P_{\pm i}^{(1)}(\omega)Z_{i\pm,s}^{(1)}(\omega)\bigg]+\sum_{i=\pm}Z_{\pm i,s}^{(1)}(\omega)Z_{i\pm,s}^{(1)}(\omega),\nonumber\\
\tilde{Z}_{\pm\mp,s}^{(2)}(\omega)&=&
\frac{1}{4\beta_0+P_{\pm\pm}^{(0)}(\omega)
-P_{\mp\mp}^{(2)}(\omega)}\bigg[P_{\pm\mp}^{(2)}(\omega)
-\left(P_{\pm\mp}^{(1)}(\omega)-P_{\pm\pm}^{(0)}(\omega)Z_{\pm\mp,s}^{(1)}(\omega)
\right)\frac{\beta_1}{\beta_0}\nonumber\\
&&-\sum_{i=\pm}P_{\pm i}^{(1)}(\omega)Z_{i\mp,s}^{(1)}(\omega)\bigg]+\sum_{i=\pm}Z_{\pm i,s}^{(1)}(\omega)Z_{i\mp,s}^{(1)}(\omega),\nonumber\\
\tilde{Z}_{\pm\mp,g}^{(2)}(\omega)&=& \tilde{Z}_{\pm\mp,s}^{(2)}(\omega)\,\frac{P_{qq}^{(0)}(\omega)
-P_{\mp\mp}^{(0)}(\omega)}{P_{qq}^{(0)}(\omega)-P_{\pm\pm}^{(0)}(\omega)}.
\end{eqnarray}

It is a well known fact that the multiplicity can be obtained from
the DGLAP evolution equations Eq.(\ref{ap}) once one is able 
to take its first Mellin moment $\omega=N-1=0$. This is not possible 
using a fixed order computation because of the presence of
singularitues at $\omega=0$ due to multiple soft emissions. 
Resummation of these divergences has been shown to be the appropriate
thing to do to avoid this problem. This has been shown a long
time ago in \cite{Mueller:1981ex} at leading logarithmic accuracy
(LL). The algebraic relations in Ref. \cite{Mueller:1982cq} show 
that the first Mellin moment of the resummed leading logarithmic 
splitting function $P_{++}^{LL}(\omega)$ can be obtained
by taking the LO leading singular term and assign an effective value to
$\omega$\,:
\begin{equation}
P_{++}^{LL}(\omega=0)=\frac{\alpha_s C_A}{\pi \omega_{eff}^{LL}}; \quad
\omega_{eff}^{LL}=2\, P_{++}^{LL}(\omega=0)=\sqrt{\frac{2C_A\alpha_s}{\pi}}
=1.382\,\sqrt{\alpha_s}
\end{equation}

\begin{figure}
\begin{minipage}[b]{6.9cm}
\centering
\includegraphics[scale=0.735]{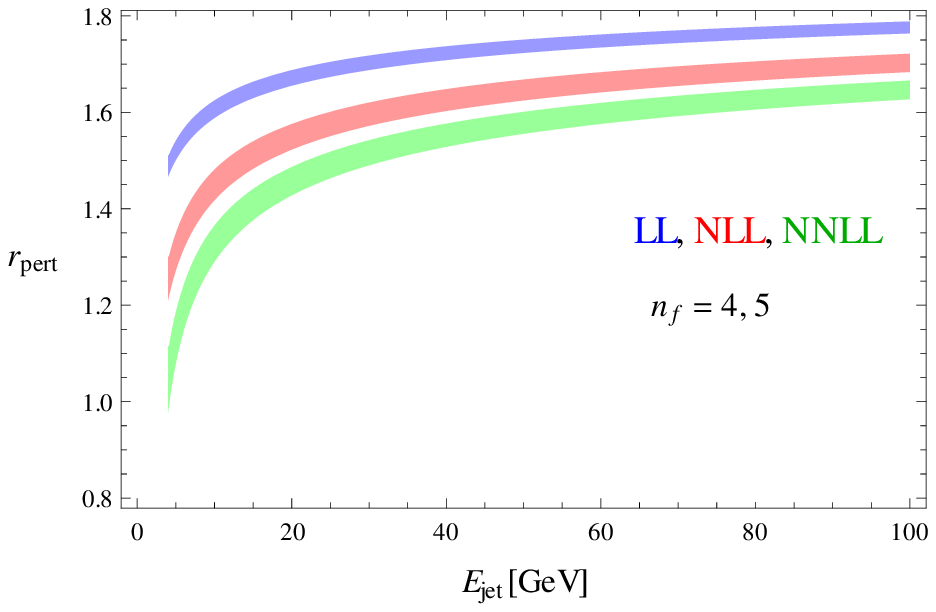}
\caption{\footnotesize{$r_{pert}^{N^kLL}(Q^2)$ with $k=0,1,2$ (blue,red,green) according to Eq.(\ref{rpert})}. The bands
correspond to $n_f=4,5$.}
\label{Fig:r_pert}
\end{minipage}
\hspace{8mm}  
\begin{minipage}[b]{6.8cm}
\centering
\includegraphics[scale=0.66]{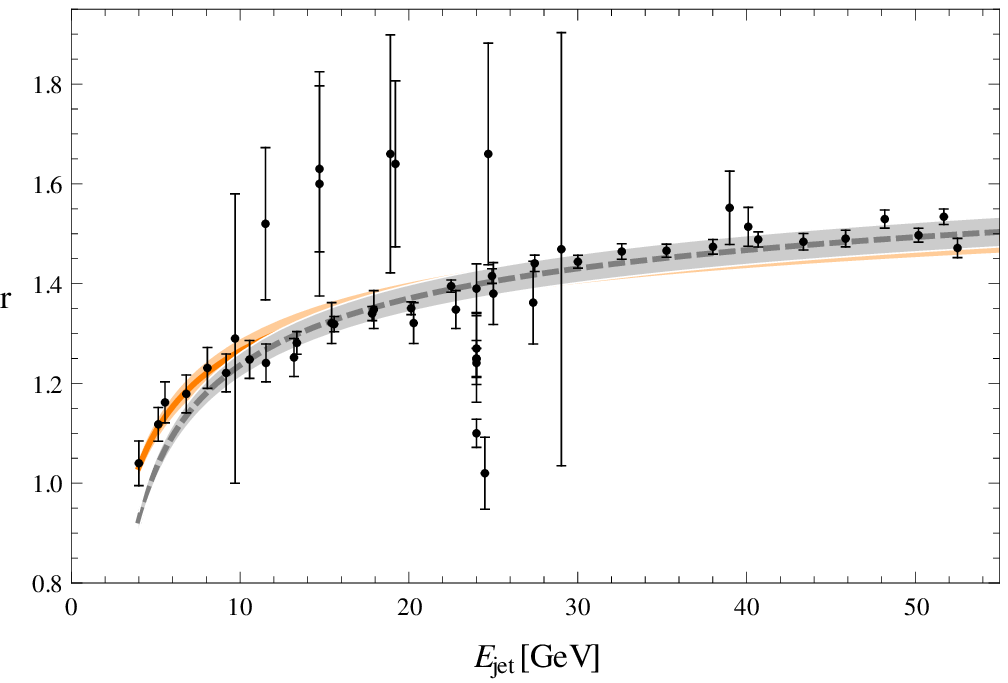}
\caption{\footnotesize{The multilicity ratio $r$} according to Eq.(\ref{rdefsimp2}) with 
$K(n_f=4)=0.9242\pm0.0186$ (orange) and  $K(n_f=5)=0.9612\pm0.0191$ (dashed grey).}
\label{Fig:K_fit}
\end{minipage}
\end{figure}

Our approach consists in adopting the same procedure also to fix $P_{++}^{NLL}(\omega=0)$
and $P_{++}^{NNLL}(\omega=0)$. The former quantity is analytically known 
\cite{Vogt:2011jv,Albino:2011cm}, while the last one is known 
up to the $16$th order \cite{Vogt:2011jv}. In this case we have obtained a numerical
estimation of the first Mellin moment of $P_{++}^{NNLL}(\omega)$ performing a numerical
extrapolation. Our result is:
\begin{equation}
\omega_{eff}^{NNLL}=1.3820\,\sqrt{\alpha_s}+
(0.0059\,n_f + 0.8754) \alpha_s+
(0.0300\,n_f + 1.0881) \alpha_s^{3/2},
\end{equation}
which is valid for $n_f=4,5$ number of active flavors. 

Neglecting the evolution of the minus component in Eq.(\ref{decomp}) and using Eq.(\ref{rlo}) we arrive at our definition of the gluon-quark multiplocity 
ratio which is given by
\begin{equation}
\label{rdefsimp2}
r^{N^k LL}(Q^2)\equiv \frac{D_g(\omega_{eff}^{N^k LL},Q^2)}
{D_s(\omega_{eff}^{N^k LL},Q^2)}=
K\,r_{pert}^{N^k LL}(Q^2),
\end{equation}
where 
\begin{equation}
\label{rpert}
r_{pert}^{N^k LL}(Q^2)=\frac{D_g^+(\omega_{eff}^{N^k LL},Q^2)}
{D_s^+(\omega_{eff}^{N^k LL},Q^2)}=-\frac{\alpha_\omega}{\epsilon_\omega}
\frac{H^+_g(\omega_{eff}^{N^k LL},Q^2)}{H^+_s(\omega_{eff}^{N^k LL},Q^2)};
\quad K=\frac{D_s^+(\omega_{eff}^{N^k LL},Q^2)}{D_s^+(\omega_{eff}^{N^k LL},Q^2)+\bar{D}_s},
\end{equation}
by use of Eq.(\ref{evolsol}).
Fig.\ref{Fig:r_pert} shows our results for $r_{pert}^{LL}(Q^2)$,$r_{pert}^{NLL}(Q^2)$
and $r_{pert}^{NNLL}(Q^2)$ for $n_f=4,5$ and Fig.\ref{Fig:K_fit} shows our $90\%$ C.L. fit
of $K$ in Eqs.(\ref{rdefsimp2},\ref{rpert}) using the NNLL result for $r_{pert}$. In
our analysis we have used the first three terms of the $\omega$ expansion for the splitting
functions and the double counted terms due to resummation have been subtracted. The running of
$\alpha_s$ has been evaluated at NNLO with $n_f=5$ and with $\alpha_s(M_Z)=0.118$.  
The data are taken from the 
summary tables of \cite{Siebel:2003zz} and references therein and from \cite{Acosta:2004js}.


{\raggedright
\begin{footnotesize}


 \bibliographystyle{DISproc}
 \bibliography{bolzoni_paolo.bib}
\end{footnotesize}
}


\end{document}